 \documentclass[USenglish,cleveref, autoref, thm-restate]{lipics-v2021}
\usepackage{cite}

\usepackage{graphicx}
\usepackage{textcomp}
\usepackage{xcolor}
\usepackage{url}
\usepackage{booktabs}

\usepackage{graphicx}
\usepackage{tikz}
\usetikzlibrary{cd}
\usepackage{thm-restate}
\usepackage{mathtools}

\nolinenumbers



\author{Ivor van der Hoog}{IT University Copenhagen,  Denmark}{ivva@itu.dk}{
https://orcid.org/0009-0006-2624-0231}{}

\author{John Iacono}{Université libre de Bruxelles, Belgium}{john@johniacono.com
}{https://orcid.org/0000-0001-8885-8172}{}

\author{Eva Rotenberg}{IT University of Copenhagen, Denmark}{erot@itu.dk}{
https://orcid.org/0000-0001-5853-7909}{}

\author{Daniel Rutschmann}{Institute of Science and Technology Austria (ISTA), Klosterneuburg, Austria}{drutschm@ista.ac.at}{
https://orcid.org/0009-0005-6838-2628}{}

\def\BibTeX{{\rm B\kern-.05em{\sc i\kern-.025em b}\kern-.08em
    T\kern-.1667em\lower.7ex\hbox{E}\kern-.125emX}}

\title{Near-Optimal Working-Set Heaps and Dijkstra on Pointer Machines}

\titlerunning{Near-optimal working-set heaps and Dijkstra on pointer machines}

\authorrunning{Ivor van der Hoog,  John Iacono, Eva Rotenberg, and Daniel Rutschmann}

\Copyright{Ivor van der Hoog,  John Iacono, Eva Rotenberg, and Daniel Rutschman}

\hideLIPIcs

\funding{\textit{Ivor van der Hoog, Eva Rotenberg, and Daniel Rutschmann} thank the VILLUM Foundation grant (VIL37507) ``Efficient Recomputations for Changeful Problems'' for supporting this work.
\textit{Daniel Rutschmann} is supported by the European Research Council (ERC) under the European Union's Horizon 2020 research and innovation programme (No.\ 101019564) \raisebox{-0.2cm}{\includegraphics[width=1.6cm]{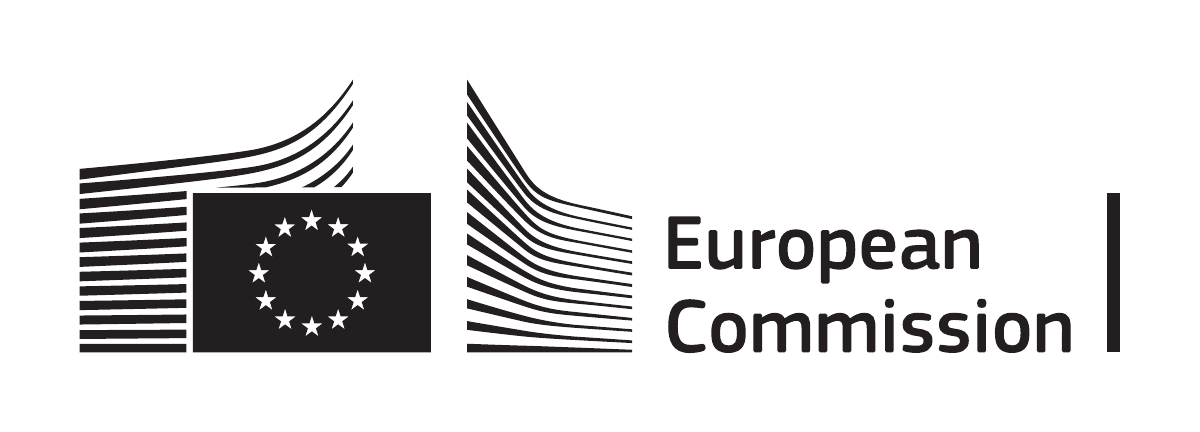}}.
\textit{John Iacono} is supported by the Fonds de la Recherche Scientifique --- FNRS.
}

\relatedversiondetails{Full Version}{https://arxiv.org/abs/2604.24134}

\begin{document}

\maketitle

\begin{abstract}
    A heap is a dynamic data structure that stores a set of labeled values under the following operations: \texttt{pop} returns the minimum value of the heap, \texttt{Push}($x_i$) pushes a new value $x_i$ onto the heap, and \texttt{DecreaseKey}($i$, $v$) decreases the value $x_i$ to $v$. 
    A working-set heap is a heap that supports the $x_i \gets$ \texttt{pop}$()$ operation in $O(\log \Gamma(x_i) )$ time where $\Gamma(x_i)$ is the size of the \emph{working set}: the number of elements that were pushed onto the heap while $x_i$ was in the heap. 
    The goal of working set heap design is to maintain the working set property while minimizing the overhead of the \texttt{Push} and \texttt{DecreaseKey} operations.
    On a word RAM, there exist working set heaps that support \texttt{Push} and \texttt{DecreaseKey} in amortized constant time. 
    In this paper, we show via a simple construction that pointer machines, one of the most general and least-assuming computational models, support working set heaps that support \texttt{Push} in amortized constant time and \texttt{DecreaseKey} in inverse-Ackermann time. A by-product of this analysis is that  Dijkstra's shortest path algorithm can be near-universally optimal on a pointer machine --- incurring only an additive $O(m \, \alpha(m))$ overhead compared to the optimal running time for distance ordering, where $m$ denotes the number of edges in the graph.
\end{abstract}
\section{Introduction}

Heap data structures are one of the fundamental data structures in theoretical computer science,
and there exist many heap implementations in the literature.
Most of the heaps in the literature are based on heap-ordered trees, i.e.~tree structures where the data stored in a node has a value that is not smaller than the value of the data stored in its parent.

At a high level, a heap is a data structure used to repeatedly extract the minimum of a dynamic set. Formally, the $\texttt{Push}$ operation adds a new data node onto the heap, and the $\texttt{pop}$ operation extracts the minimum of the current heap. 
Heaps typically also support various other operations: \texttt{MakeHeap} creates a new heap and the \texttt{Peek} operation yields the minimum value in the heap without removing it. 
Another common and useful operation is the \texttt{DecreaseKey} operation which selects a current data node in the heap and replaces the data with a smaller value. The combination of \texttt{DecreaseKey} and \texttt{Pop} then allows for arbitrary deletions from the heap.

We are interested in implementing heaps on a pointer machine --- one of the most general models of computation, and one of the least-assuming generalizations of the RAM model used to analyze data structures.
The classical pointer machine solution is a heap-ordered tree, which achieves logarithmic time for all aforementioned heap operations. Early examples are the implicit binary heaps of Williams~\cite{WilliamsHeapsort}, the leftist heaps of Crane~\cite{crane1972linear} and Knuth~\cite{DBLP:books/aw/Knuth68}, and the binomial heaps of Vuillemin~\cite{Vuillemin2078}. Fredman and Tarjan introduced Fibonacci heaps~\cite{fredman1987fibonacci} which achieved $O(1)$ amortized time for all the  above operations except for \texttt{Pop}, which requires $O(\log n)$ amortized time where $n$ is the number of data nodes in the heap. 
Fibonacci heaps had an immediate and famous application in Dijkstra's algorithm for single-source shortest paths~\cite{Dijkstra1959}.
If Dijkstra's algorithm is run on a graph with $n$ nodes and $m$ edges then, using a Fibonacci heap, the algorithm takes $\Theta(m + n \log n)$ time, which is worst-case optimal.

Since the introduction of Fibonacci heaps, a variety of other heaps with fast \texttt{DecreaseKey} have been introduced (see \cite{Chan2013} for a full survey of priority queues):
The Pairing heap~\cite{Fredman1986Pairing} is a self-adjusting structure modeled after splay trees, whose asymptotic amortized \texttt{DecreaseKey} running time is unknown.
Only an upper bound of $O(2^{\sqrt{\log \log n}})$ \cite{Pettie2005Towards} and a lower bound of $\Omega(\log \log n)$ \cite{Fredman1999Pairing} are known.
The strict Fibonacci heap has worst-case rather than amortized running times \cite{Brodan2025Strict}.
Chan's quake heaps were invented to be a simpler alternative to Fibonacci heaps with $O(1)$ amortized time \texttt{DecreaseKey}~\cite{Chan2013}.
Rank-pairing heaps \cite{Haeupler2011Rank} are another alternative to Fibonacci heaps with $O(1)$ amortized time \texttt{DecreaseKey} but their structure is closer to the pairing heap.
Violation heaps~\cite{Elmasry2010Violation}, 
Hollow Heaps~\cite{Hansen2017Hallow}, and Fibonacci Heaps revisited~\cite{KaplanTarjanZwick2014} are other variants with $O(1)$ amortized time \texttt{DecreaseKey}.
Slim and smooth heaps \cite{hartmann2021Analysis} have $O(\log \log n)$ \texttt{DecreaseKey}, but achieve constant indegree.


%
\subparagraph{Adaptive bounds.}
 An adaptive heap supports the \texttt{Pop} operation in $o(\log n)$ amortized time under specific circumstances,
 while the \texttt{Push} operation takes $O(1)$ amortized time.
 Most existing heaps with adaptive bounds work without supporting the \texttt{DecreaseKey} operation:
 Iacono~\cite{iacono2000improved} shows that, on a pairing heap, $x \gets $\texttt{Pop}$()$ takes $O(1+ \log \Gamma(x))$ time
 where $\Gamma(x)$ is the number of data nodes that were pushed while $x$ was on the heap.
 Elmasry~\cite{elmasry_priority_2006} designs a heap on which $x \gets $\texttt{Pop}$()$ takes $O(1 + \log \Gamma'(x))$ time
 where $\Gamma'(x)$ is the number of data nodes that were pushed while $x$ was on the heap
 and remain in the heap until $x$ is popped.
 Elmasry, Farzan and Iacono~\cite{elmasry_priority_2012} improve this to $O(\min(\log \Gamma'(x), \log \ell))$ time
 where $\Gamma'(x)$ is defined in the same way and $\ell$ is the number of data nodes that were pushed \emph{before}
 $x$ was pushed and that remain in the heap until $x$ is popped.
 Rutschmann~\cite{rutschmann2026sorting} designs a heap with the unified bound: $x \gets $\texttt{Pop}$()$ takes $O(\min_{y} (1 + \log(a(x, y)) + \log(b(x, y)))$ time
 where $y$ can be any data node that was already popped, $a(x, y)$ is the number of push operations
 between \texttt{Push}($x$) and \texttt{Push}$(y)$, and $b(x, y)$ is the number of \texttt{Pop} operations between $x \gets$\texttt{Pop}$()$ and $y \gets$\texttt{Pop}$()$.
 Kozma and Saranurak~\cite{kozma_smooth_2018} introduce the Smooth heap,
 which is conjectured to be instance optimal --
 if true, this implies that the smooth heap has all the above bounds.

 We are aware of two adaptive heaps that support \texttt{DecreaseKey}.
 Both  implement the \emph{working set property}: \texttt{Push} and \texttt{Peek} take (amortized) constant time, and \texttt{Pop} takes $O(1+ \log \Gamma(x))$ time, 
 where $\Gamma(x)$ is the number of data nodes that were pushed while $x$ was on the heap. 
 Haeupler, Hlad{\'\i}k,  Rozho{\v{n}}, Tarjan, and  T{\v{e}}tek~\cite{Haeupler24} present a word RAM implementation of a working set heap that moreover supports \texttt{DecreaseKey} in amortized constant time. 
They show that using this heap, Dijkstra's algorithm is universally optimal on the word RAM (see  Corollary~\ref{cor:dijkstra}). 
Van der Hoog, Rotenberg, and Rutschmann~\cite{hoog2025SimplerDijkstra} simplify their result.

\subparagraph{RAM Model.}
The two adaptive heaps that support \texttt{DecreaseKey} are implemented not on a pointer machine, but on a word RAM.
The word RAM model is very powerful, and in this model it is frequently possible to improve upon the best pointer-machine model results, typically by double-log factors. Fundamentally, from any given node in a pointer machine, one can only access a constant number of nodes. In contrast, in the RAM model one can access an arbitrary element of an array in constant time.
Furthermore, if on a RAM the data is assumed to have a certain representation, such as word-sized integers rather than simply comparison-based objects, the world of what is commonly known as \emph{bit-tricks} becomes available for data structure implementations. These were introduced by Fredman and Willard with Fusion trees \cite{fredman1993surpassing} and extended to the heap setting with Atomic Heaps \cite{Fredman2994Trans}; further improvements were made by Thorup in relation to the circuit complexity~\cite{DBLP:conf/soda/Thorup03a}.

\subparagraph{Pointer machine data structures.}
In this paper, we find it imperative to establish what the best pointer machine model solution is for any problem.
This is for a number of reasons:
\begin{itemize}
    \item First, we note that pointer machines are the more general model of computation. RAM models often allow for faster running times than pointer machine solutions, as RAM tricks  often squeeze out $\operatorname{loglog}$ or similar factors from the best pointer machine result. In the interest of generality and purity of the model of computation, we therefore want to see what overhead is obtained by relying on a pointer machine rather than a Word RAM.
    \item Secondly, improvements in the pointer machine model often lead to better RAM model algorithms as efficient solutions in this model require fundamental algorithmic efficiency.
    \item Thirdly, many of the techniques used to speed up RAM algorithms are invoked to shave off log-factors in manners that do not always translate to practice. In practice, such bit-trick techniques require careful packing of multiple items into machine words to exploit word-level parallelism which in itself introduces practical overhead. 
    Pointer machine solutions, such as ours, are in comparison typically simple and implementable.
    \item Fourthly,  the pointer machine model lends itself more naturally to lower bounds. 
    For example, the union-find problem has a well-known inverse-Ackermann lower bound that is derived in the pointer machine model. Achieving strong upper bounds in the pointer machine model can indicate a potential for lower bounds that cannot exist on a RAM. 
\end{itemize}
This makes the pointer machine model the natural model to analyze data structures.

\subparagraph{Our results.}
Our goal is to find a pointer machine heap with the working set property that supports the \texttt{DecreaseKey} operation with small overhead.
We aim to provide a general and simple, all-purpose pointer machine solution. 
Previous works with adaptive heap bounds either (1) disallow \texttt{DecreaseKey}, (2) support \texttt{DecreaseKey} in logarithmic or doubly logarithmic time, or (3) use the sledgehammer of RAM tricks to solve small-size problems in constant time and eliminate iterated-log terms from otherwise inefficient structures.  

We present a pointer-model heap with constant-time \texttt{Push}, working-set \texttt{Pop}, and inverse-Ackermann time \texttt{DecreaseKey}. It is based on a simple, recursive, and elegant construction using Fibonacci heaps or other optimal pointer machine model non-adaptive heaps as building blocks.
While our implementation is not necessarily simpler than the RAM implementations~\cite{Haeupler24, Hoog2025Simpler},
we consider our solution considerably simpler compared to other adaptive pointer machine heaps~\cite{iacono2000improved, elmasry_priority_2006, elmasry_priority_2012} and in addition, we support the \texttt{DecreaseKey} operation.

The inverse-Ackermann is a stubborn bottleneck in our approach, and all our attempts at removing this overhead have failed. This leaves open the question of whether the inverse-Ackermann is \emph{necessary} in the pointer machine model. For example, is a reduction possible to other problems with inverse-Ackermann running times such as union-find?

\section{Preliminaries}

We  implement working-set heaps on a pointer machine. 
Knuth~\cite{DBLP:books/aw/Knuth68} gave one of the earliest formal definitions of a pointer machine model, which on a high level is a directed graph where nodes can store data values and nodes have pointers to other nodes. Tarjan~\cite{TARJAN1979110} refined the definition, formalizing the operations supported by a pointer machine.
In this paper, we will use a more general version of the pointer machine by Tarjan~\cite{TARJAN1979110}. 
In particular, our model of computation only requires a subset of the operations supported by Tarjan as we have no need to modify our data (only our data structure). 
For algorithmic convenience, we assume that each node in the pointer machine can store $O(1)$ pointers.
In contrast, Tarjan~\cite{TARJAN1979110} uses only one pointer, but that pointer can point to specific memory cells called \emph{pointer fields} that record any additional pointers. These assumptions are asymptotically equivalent.  

\subparagraph{Formally defining a pointer machine data structure.}
A pointer machine consists of \emph{nodes} and \emph{pointers}.
There are two types of nodes. A \emph{data node} stores any type of data (integers, logical values, strings, real numbers, whatsoever). 
A \emph{pointer node} stores a number of constantly many bits. 
Each node can have up to constantly many pointers, where a pointer is a directed edge from one node to another. 
The \emph{dynamic input} is a set of data nodes, where the current input size $n$ denotes the number of data nodes.
An \emph{update} adds, removes, or changes the data of a data node.
A \emph{dynamic data structure} maintains an arbitrarily-large set of pointer nodes and the pointer nodes of the data nodes.
One pointer node is called the \emph{root} and we maintain the invariant that any node can be reached by a directed path from the root. The \emph{size} of the data structure is the total number of nodes it has. 

Any update or query to the data structure runs a \emph{program} with a \emph{working memory}: an array of constant size where each cell points to a node.  It executes a sequence of \emph{instructions}:
\begin{itemize}
    \item A \emph{comparison} takes as input two nodes in working memory of the same type (pointer or data nodes) and adds to working memory a pointer node with the Boolean outcome.  
    \item An \emph{atomic operation} takes as input three nodes in working memory. It applies the operator to the latter two and replaces the first node with its output. 
    \item A \emph{copy} updates the working memory by replacing one node by the copy of another.   
    \item A \emph{traversal} updates the working memory by replacing one node by a node it points to. 
        \item A \emph{halt} instruction ends the computation and the output is the current working memory.
\end{itemize}
By this definition, input values can be compared but not manipulated. This makes the model very general. 
The \emph{running time} is the total number of instructions until a \emph{halt} instruction.

Worst-case running time bounds imply amortized bounds.

\subparagraph{Defining heaps.}
Having defined our operations, we abstractly define a \emph{heap} data structure. 

\begin{definition}
A \emph{heap} data structure stores a set of data nodes with data values in some linearly ordered universe. Let $n$ denote the current number of data nodes. 
A heap supports the following operations:
\begin{itemize}
    \item \texttt{MakeHeap}() creates a new data structure and returns a pointer $r$ to its root.
    \item \texttt{Push}($r$, $x_j$) Given a pointer $r$ to a heap's root, and a data node $x_j$, add $x_j$ to the heap.
    \item \texttt{Peek}($r$) Given a pointer to the root $r$ of a heap, output the minimum value on that heap. 
    \item \texttt{Pop}($r$) Given a pointer to the root $r$ of a heap, output and remove its minimum value. 
    \item \texttt{DecreaseKey}($r$, $x_j$, $v$) Given a pointer $r$ to the root of a heap, a data node $x_j$ in that heap, and new value $v$ smaller than the current value of $x_j$, replace the value of $x_j$ with $v$.     
    \item (optional) \texttt{Meld}($r$, $r'$) Given two pointers $r$ and $r'$ to two distinct heap data structures returns a pointer to one heap containing all elements. 
\end{itemize}
\end{definition}
The interface of the \texttt{DecreaseKey} operation has a subtle complication:
It requires not only a pointer to the data node, but also
a pointer to the root of the data structure. Matching this interface is difficult when we maintain several heaps subject to the Meld operation~\cite{Kaplan2022Meld}. 
Indeed, if our memory contains heaps $H_1, \ldots, H_k$ and we wish to perform \texttt{DecreaseKey} on some element $x_j \in H_i$ then the \texttt{DecreaseKey} interface demands a pointer to the root of $H_i$ (in addition to the pointer to $x_j)$.
Explicitly storing a pointer from each data node $x_j$ to the root of its heap $H_i$
would solve this issue. 
However these pointers cannot be maintained efficiently under the Meld operation (as linearly many data nodes may need a new pointer).

\subparagraph{Fibonacci heaps.}
Fredman and Tarjan~\cite{fredman1987fibonacci} present a pointer machine implementation of a Fibonacci heap, which supports the \texttt{MakeHeap}, \texttt{Peek}, \texttt{Push}, and \texttt{DecreaseKey} operations in amortized constant time and the \texttt{Pop} operation in amortized logarithmic time. 
In addition, their heap supports the Meld operation in worst-case constant time. However, as noted earlier, to support Meld, the interface of the \texttt{DecreaseKey} operation requires an additional pointer.
Brodal, Lagogiannis and Tarjan~\cite{Brodan2025Strict} achieve the same bounds with worst-case time instead of amortized time, with the same caveat on combining \texttt{Meld} and \texttt{DecreaseKey} operations. 

\subparagraph{Working set heaps.}
Iacono and Fredman~\cite{iacono2001distribution} introduce several notions of \emph{working set size}, some of which are asymptotically equivalent. Each notion assigns to every heap element $x_j \in H$ a value $\Gamma(x_j)$, called its \emph{working set size}. The objective is to design heap data structures that support $\texttt{Pop}$ in time $O(1 + \log \Gamma(x_j))$, where $x_j$ is the element returned.
In this paper, we focus on a specific definition from \cite{iacono2001distribution} that is used in recent work~\cite{Haeupler24, hoog2025SimplerDijkstra}. To distinguish it from the variants in~\cite{iacono2001distribution}, we instead refer to this quantity as the element’s \emph{age}:

\begin{definition}
    Let $H$ be a heap of size $n$. For any data node $x_j \in H$ we define its \emph{age} $\Gamma(x_j)$ as the number of elements that were pushed onto the heap after $x_j$ (including $x_j$ itself). A \emph{working set} heap supports the \texttt{Push} and \texttt{Peek} in $O(1)$ time, \texttt{Decreasekey} in $O(f(n))$ time, and the \texttt{Pop} operation in $O(1 + \log \Gamma(x_j))$ time where $x_j$ is the returned element. 
\end{definition}

The goal of designing an efficient working set heap is to provide an implementation where the  `overhead' $f(n)$ is as low as possible.
Iacono~\cite{iacono2000improved} shows that a double pairing heap can support the \texttt{Push}, \texttt{Peek} and \texttt{Meld} operations in amortized constant time and the \texttt{Pop} operation in amortized $O(1 + \log \Gamma(x_j))$ time. However, this analysis does not support \texttt{DecreaseKey}.
Haeupler, Hlad{\'\i}k,  Rozho{\v{n}}, Tarjan, and  T{\v{e}}tek~\cite{Haeupler24} present a word RAM implementation of a working set heap with no overhead (i.e. $f(n) = 1$) where all running times are amortized. 
They show that using this heap, Dijkstra's algorithm is universally optimal on the word RAM (we forward reference Corollary~\ref{cor:dijkstra} for the formal statement). 

Van der Hoog, Rotenberg, and Rutschmann~\cite{hoog2025SimplerDijkstra} simplify their analysis and show a simpler word RAM implementation that achieves amortized bounds. Both solutions crucially rely on bit-tricks. We argue later that replacing the bit-tricks with pointer machine operations naively incurs an overhead of $f(n) = \log \log \log n$. 
We show that it is possible to achieve $f(n) = \alpha(n)$ where $\alpha(n)$ is the inverse-Ackermann function.  Note that we do not support \texttt{Meld} and that it is even unclear how the ages of elements would be defined under \texttt{Meld}.

\begin{restatable}{theorem}{main} \label{theo:main}
    A heap can be implemented on a pointer machine that supports the following operations, where $n$ denotes the heap size:
    \begin{itemize}
        \item \texttt{MakeHeap} in worst-case $O(1)$ time,
        \item \texttt{DecreaseKey} in amortized $O(\alpha(n))$ (inverse-Ackermann) time, 
        \item \texttt{Push} in amortized $O(1)$ time, 
        \item \texttt{Peek} in worst-case $O(1)$ time, and
        \item  \texttt{Pop} in worst-case $O(1 + \log \Gamma(x_j))$ time.
    \end{itemize}
    All worst-case operations build no amortized potential. I.e., invoking them does not affect the amortized analysis of the other operations. 
\end{restatable}

\begin{corollary}[Direct consequence of~\cite{Haeupler24, hoog2025SimplerDijkstra}]
\label{cor:dijkstra}
Fix any (directed) graph $G$ with $n$ vertices and $m$ edges, where one vertex $s$ is denoted as the source. 
Denote for a fixed single-source shortest path (SSSP) algorithm $\mathcal{A}$ by $U(\mathcal{A}, G)$ the maximum running time required by $\mathcal{A}$ to sort all vertices by their distance from $s$ (the maximum is taken over all positive edge-weightings of $G$). 
Let $\mathcal{U}$ denote the minimum over all SSSP algorithms $\mathcal{A}$ of $U(\mathcal{A}, G)$.
Then Dijkstra's algorithm equipped with the heap from Theorem~\ref{theo:main} achieves a running time of $O(\alpha(n) \cdot m + \mathcal{U})$.    
\end{corollary}
 
\section{A working set heap}

The basis of a working set heap, as shown in~\cite{Iacono2001Alternatives}, is an exponential bucketing scheme based on the \emph{age} of data nodes in the heap. 
Consider all elements $x_j \in H$ ordered by their age. 
This bucketing scheme creates an ordered doubly linked list of pointer nodes (called \emph{buckets}) $H_i$. One may choose whether these grow exponentially or doubly exponentially in size.
Each bucket $H_i$ stores a Fibonacci heap, such that the buckets together partition $H$. Moreover, for all $i$, the data in $H_i$ has a lower age than the data in $H_{i+1}$. 
The goal is to ensure that for any $x_j \in H_{i}$, the logarithm of its age ($\Theta(\log \Gamma(x_j))$) lies in $\Omega(\log |H_{i}|)$. 
Crucially, for each bucket $H_i$ the \texttt{Pop} operation (returning $x_j$) then runs in $O(\log \Gamma(x_j))$ time as Fibonacci heaps support \texttt{Pop} in time $O(\log N)$ where $N$ denotes the size of the Fibonacci heap. 

\subparagraph{Maintaining the buckets and pointer machines.}
On a pointer machine these buckets $H_1, \ldots, H_k$ are pointer nodes that form a doubly linked list where for each bucket $H_i$, its third pointer points to the root of the Fibonacci heap $T(H_i)$. 
The high-level data structure can easily be maintained in amortized constant time via a strategy that greedily merges adjacent buckets~\cite{Haeupler24,hoog2025SimplerDijkstra}.
However, two key challenges remain:
\begin{enumerate}
    \item Firstly, to support \texttt{DecreaseKey}$(r, x_j, v)$ on an element $x_j \in H$, we need to invoke \texttt{DecreaseKey}$(r', x_j, v)$ on the Fibonacci heap $T(H_i)$ where $x_j \in H_i$ and where $r'$ is a pointer to the root of $T(H_i)$. 
    However, we already remarked that it is nontrivial to obtain $r'$. 
    We want to meld Fibonacci heaps in constant time, so we cannot explicitly maintain a pointer from each element $x_j$ to the root of the melded heap. Rather, we need to spend time during a \texttt{DecreaseKey} operation to determine the current heap that $x_j$ is in.
    \item Secondly, to support \texttt{Peek} and \texttt{Pop} on the heap $H$, we need to maintain the index $i$ such that $H_i$ contains the minimum element.
\end{enumerate}
These issues are essentially bypassed on a word RAM. 
For the first problem, the buckets can be stored in an array. Consider an element $x_j$ where $i := \lfloor \log \Gamma(x_j) \rfloor$. These bucketing schemes ensure that $x_j$ is in some $H_k$ with $k = i \pm O(1)$. 
Let $x_j$ be the $j$'th element that was pushed onto $H$. If we at all times maintain a pointer to the last element $x_k$ that was pushed onto the heap, then we can compute for any element $x_j$ its age via the formula $k-j+1$. 
As a result, one can use the age of an element $x_j$ to compute the index of bucket that contains $x_j$, and obtain the root of the corresponding Fibonacci heap.
For the second problem, one maintains the minimum of at most $\log n$ values. By representing these values as a bitstring (where a $0$ at index $i$ denotes that the value in $H_i$ is smaller than in $H_r$ for all $r > i$) and using constant-time bitstring manipulation one can maintain the index $i$ such that $H_i$ stores the minimum of $H$ at no additional overhead. 

\subparagraph{A naive pointer machine implementation.}
On a pointer machine, existing solutions to these two problems yield a working set heap with \texttt{DecreaseKey} overhead $f(n) = \log \log \log n$:
The union-find data structure~\cite{tarjan1975efficiency} is a pointer machine data structure that maintains, for a partition of data nodes into sets, a representative node from each set.
It supports the \texttt{MakeSet} operation in amortized $O(1)$ time, which creates a new singleton set;
the \texttt{Link} operation in amortized $O(1)$ time, which takes as input two representatives of different sets,
replaces these sets by their union, and returns a new representative;
and the \texttt{Find} operation in amortized $O(\alpha(n))$ time, which takes a data node and returns the representative of its set.
Here, $n$ is the total number of \texttt{MakeSet} operations performed.
By applying \texttt{Link} whenever two Fibonacci trees get melded, and storing pointers between the Fibonacci tree and the representative,
one can use \texttt{Find} to return for each element $x_j$ the root of its Fibonacci tree in $O(\alpha(n))$ time. 
For the second problem, note that when using doubly exponentially growing buckets there are $O(\log \log n)$ such minima. Maintaining these in a separate binary heap incurs an additive $O(\log \log \log n)$ overhead onto \texttt{Push}, \texttt{DecreaseKey}, and \texttt{Pop}. If we use a suitably biased binary heap, we only incur this $O(\log \log \log n)$ overhead on \texttt{DecreaseKey}. The challenge in this paper is to maintain a working set heap that supports the \texttt{Push} and \texttt{Peek} operation in amortized $O(1)$ time and \texttt{DecreaseKey} in amortized $O(\alpha(n))$ time.

\subsection{Our pointer machine implementation.}

In Section~\ref{sect:top_level}, we design an auxiliary data structure that supports the following:
\begin{restatable}{theorem}{thmTop} \label{theo:top_level}
\textbf(Index structure)
    We can maintain a collection of $m$ data nodes $A = \langle A[1], A[2], \ldots, A[m] \rangle$ subject to the following operations:
    \begin{itemize}
    \item $p_m\gets\texttt{Push}(A[m])$: increases the size of the structure by adding a new last element $A[m]$ and returns a pointer $p_m$ to be used by the other operations to change $A[m]$. This runs in worst-case $O(m)$ time.
    \item $p\gets\texttt{Peek}(A)$: returns a pointer $p$ to the minimum value in $A$, in worst-case $O(1)$ time.
\item $\texttt{DecreaseKey}(p_i, v)$: decreases the value of $A[i]$ to $v$ in worst-case $O(\alpha(i))$ time.
\item $\texttt{ChangeKey}(p_i, v)$: changes the value of $A[i]$ to $v$ in worst-case $O(i)$ time.
    \end{itemize}
\end{restatable}

\noindent
We now describe our main data structure, using the index data structure as a black box. 
Our data structure will maintain a partition of the current stored data into sets $H_1, H_2, \ldots H_m$, which we refer to as \emph{buckets} with the following three invariants:
\begin{itemize}
    \item Each bucket $H_i$ either corresponds to a strict Fibonacci heap $F_i$ containing $H_i$ (as strict Fibonacci heaps exist in the pointer machine model), or, is \emph{vacant}. The \emph{vacancy invariant} requires that only buckets $H_i$ with odd $i$ may be vacant. We use the word vacant to mean there is no structure $F_i$, to distinguish from \emph{empty}, which refers to a Fibonacci heap $F_i$ that is present but stores no data.
    \item The \emph{size invariant} requires the maximum size of $H_i$ to be $2^{\lfloor i/2 \rfloor}$.
    \item The age of all items in bucket $i$ is at least the sum of the maximum sizes of the non-vacant smaller buckets. Formally, the \emph{age invariant} states that:
$$
\text{If $x \in H_i$ then }\Gamma(x) \geq
\sum_{\mathclap{\substack{j<i\\j\text{ non-vacant}}}} 2^{\lfloor j/2 \rfloor}
$$
Since all $H_i$ for even $i$ are non-vacant, for $x \in H_i$, this yields $\Gamma(x) \geq 2^{i/2-2}$.
\end{itemize}
The above properties are designed so that (a) swapping $H_{2i}$ and $H_{2i+1}$ where one of them is vacant and (b) melding nonvacant $H_{2i}$ and $H_{2i+1}$ and placing the result into the vacant $H_{2i+2}$ (leaving $H_{2i}$ and $H_{2i+1}$ vacant) both maintain the size and age invariants (but not the vacancy invariant).
The minimum value in each $H_i$ is stored as $A[i]$ with pointer $p_i$ in the index structure (Theorem~\ref{theo:top_level}). If $H_i$ is vacant or empty then we store $\infty$ instead. 

Additionally, each data node is stored in a union-find structure which maintains the current partition into the buckets. We use $r_i$ to denote the union-find data structure's representative of $H_i$. I.e., given a pointer $\rho$ to a data node $x$ in $H_i$, $\texttt{Find}(\rho)$ returns $r_i$.

To be clear: our data structure stores three things for each bucket $H_i$, the Fibonacci heap pointer $F_i$, the pointer $p_i$ into the index structure, and the representative from the union-find structure $r_i$. Pointers to and between all of these, for a given $i$, are all stored in a linked list ordered by $i$.
The general idea is to use Bentley-Saxe~\cite{BENTLEY1980301} inspired insertions on the exponentially-sized $H_i$'s, made faster through the constant-time \texttt{Meld} operation of Fibonacci heaps. This is augmented by the union-find structure that allows us to find out efficiently which $H_i$ a given data node is in, as well as the indexing structure whose primary role is to quickly determine which $H_i$ currently stores the minimum element.

We now describe how to implement each heap operation. We define here one recurrent subroutine $\texttt{Update}(p_i)$ which updates an out-of-date value in the index structure
by obtaining the minimum value of $F_i$ via $m_i=\texttt{Peek}(F_i)$ and then invokes $\texttt{ChangeKey}(p_i, m_i)$ in the indexing data structure  to update $A[i]$. This takes worst-case time $O(i)$.

\begin{description}
    \item[\texttt{DecreaseKey:}] 
    To perform $\texttt{DecreaseKey}(\rho, v)$, given the pointer $\rho$ to the data node $x$ to decrease, we first invoke $r_i=\texttt{Find}(\rho)$ on our union-find data structure in amortized $O(\alpha(m))$ time to obtain the representative $r_i$ such that $x \in H_i$. With $r_i$ we also have $F_i$ and $p_i$. Since we now know which Fibonacci heap $x$ belongs to, we can invoke $\texttt{DecreaseKey}(F_i, \rho, v)$ on the strict Fibonacci heap $F_i$. Next, let $y=\texttt{Peek}(F_i)$ be the possibly new minimum of $F_i$. Finally, we need to update the index structure, this is done by invoking $\texttt{DecreaseKey}(p_i, y)$ on the index structure in $O(\alpha(i)) \subseteq O(\alpha(m))$ worst-case time.
    For now, we assume that $m \in O(n)$, so that \texttt{DecreaseKey} takes $O(\alpha(n))$ amortized time.
    \item[\texttt{Push:}] 
    We first ensure that $H_0$ is vacant (temporarily breaking the vacancy invariant) by the following recursive process: To ensure $H_{2i}$ is vacant we do one of three things. If $H_{2i}$ is not vacant but $H_{2i+1}$ is vacant, we move $H_{2i}$ to $H_{2i+1}$; if both $H_{2i}$ and $H_{2i+1}$ are vacant, we recursively ensure $H_{2i+2}$ is vacant and meld $H_{2i}$ and $H_{2i+1}$ and place the result in the now-vacant $H_{2i+2}$. 

    Given that $H_0$ is now empty, we place $x$ into a new Fibonacci heap $F_0$, create a new set with representative $r_0$ in the union-find structure, and call $\texttt{Update}(p_0)$ to update the index structure.

Moving $H_{2i}$ to $H_{2i+1}$ is just a matter of swapping pointers: $F_{2i}$ and $F_{2i+1}$;  $r_{2i}$ and $r_{2i+1}$ and the running of $\texttt{Update}(p_{2i})$ and $\texttt{Update}(p_{2i+1})$.
Melding $H_{2i}$ with $H_{2i+1}$ to yield $H_{2i+2}$ involves melding the heaps via $F_{2i+2}=\texttt{Meld}(F_{2i},F_{2i+1})$ as well as melding the union-find structure via $r_{2i+2}=\texttt{Link}(r_{2i},r_{2i+1})$. Finally the index structure is updated with calls to $\texttt{Update}$ on $p_{2i}, p_{2i+1}$ and $p_{2i+2}$.
In the special case where we are using a $H_i$ for the first time, it needs to be added to the index structure via $\texttt{Push}$.

The worst-case running time is $O(\ell^2)$ where $\ell$ is the largest value where $H_\ell$ was changed in this operation. This can be bounded by $O(1)$ amortized time by giving each non-vacant bucket $H_i$ a potential of $\Theta(i^2)$ and by applying the textbook binary counter analysis in~\cite[Chapter 17]{CLRS2009}. Note that this does not affect other operations as no other operations change whether or not a bucket is vacant.
    
    \item[\texttt{Peek:}] 
    Call $p_i=\texttt{Peek}()$ from the index structure and return the value of $A[i]$, which can be obtained from $p_i$.
    \item[\texttt{Pop:}] 
    Call $p_i=\texttt{Peek}()$ on the index structure. With $p_i$, we obtain $F_i$.
We then invoke $x=\texttt{Pop}(F_i)$ on that heap in worst-case $O(\log |H_i|) = O(i)$ time. Note that after the pop, $F_i$ may be empty, but we do not change it to vacant. Finally, invoke $\texttt{Update}(p_i, m)$ to update index structure
in $O(i)$ time, before returning $x$.
Since for each element $x$ in $H_i$, $\Gamma(x)\geq 2^{i/2-2}$, the worst-case running time of $O(i)$ can be expressed as $O(1 + \log \Gamma(x))$. 
\end{description}
Let us now address the $m \in O(n)$ assumption required by \texttt{DecreaseKey}: At the start of \texttt{DecreaseKey}, we check whether $m > n$. If so, then we rebuild the whole data structure from scratch. By the textbook analysis for dynamic arrays~\cite{CLRS2009}, this incurs $O(1)$ amortized time.
\\

\noindent
Thus, once we have proven \cref{theo:top_level}, we obtain:

\main*

\section{A Fast Index Structure} \label{sect:top_level}

We prove that we can maintain the indexing data structure (Thm.~\ref{theo:top_level}) via the following lemma:


\begin{lemma} \label{lemm:tree}
    For every integer $m$, there is a rooted tree with $m$ leaves that can be computed in $O(m)$ time and that has the following properties:
    \begin{enumerate}
        \item The $i$-th leaf has a depth of $O(\alpha(i))$
        \item Every node on the root-to-leaf path of the $i$-th leaf has $O(\sqrt{i})$ children.
    \end{enumerate}
 
\end{lemma}

\subparagraph{Showing that Lemma~\ref{lemm:tree} implies \cref{theo:top_level}.}
We first show that \cref{lemm:tree} implies \cref{theo:top_level} on a pointer machine. 
The data structure uses the tree $T$ from \cref{lemm:tree} as follows:
The $i$-th leaf stores the value $A[i]$.
Every node stores a pointer to its parent, and to its first child.
We store the children of every node in a doubly linked list.
Every node stores a pointer to the leaf that contains the minimum value in its subtree.
And finally, we store a global pointer to the root of the tree. As a result, all nodes have $O(1)$ outgoing pointers.
Given the edges of the tree, all of this can be computed in $O(m)$ time.
We use this tree to implement the operations specified in \cref{theo:top_level} as follows:  

\begin{itemize}
    \item \texttt{Peek}$(A)$: since the root node stores a pointer to the minimum values in the whole tree,
we can return this value in $O(1)$ time.
\item \texttt{DecreaseKey}$(A[i], v)$: We update the value in the $i$-th leaf. Then, we walk up the root-to-leaf path and update the subtree minimum pointer
if $A[i]$ has become the new subtree minimum. Since the $i$-th leaf has depth $O(\alpha(i))$, this takes $O(\alpha(i))$ time in total.
\item 
\texttt{ChangeKey}$(A[i], v)$ : We update the value in the $i$-th leaf. Then, we walk up the root-to-leaf path and recompute the subtree minimum pointers
from scratch by looking at the subtree minimum pointers of all children. On the root-to-leaf path,
there are $O(\alpha(i))$ nodes with $O(\sqrt{i})$ children each,
so this takes $O(\sqrt{i}\, \alpha(i)) \subseteq O(i)$ time in total.
\end{itemize}

\noindent
Thus, \cref{theo:top_level} follows from \cref{lemm:tree}.

\subsection{Proving \cref{lemm:tree}}

We show \cref{lemm:tree} via the following weaker helper lemma:

\begin{lemma} \label{lemm:tree_weak}
     For every integer $m$, there is a rooted tree with $m$ leaves, that can be computed in $O(m)$ time, that has the following properties:
    \begin{enumerate}
        \item The $i$-th leaf has a depth of $O(\alpha(i))$
        \item Every node on the root-to-leaf path of the $i$-th leaf has $O(i)$ children.
    \end{enumerate}
\end{lemma}
From \cref{lemm:tree_weak}, we can obtain \cref{lemm:tree} via a subdivision step:
For every internal node $u$ with $k \ge 4$ children, we add $\lceil\sqrt{k}\rceil$ new nodes that each have $u$ as their parent and
$\Theta(\sqrt{k})$ of the (former) children of $u$ as their children.
This at most doubles the depth of every leaf,
and reduces the number of children by a square root.
 What remains is to show \cref{lemm:tree_weak}.

\begin{definition} \label{def:up_arrow}
    \emph{Knuth's up-arrow notation} for integers $a, b, k \ge 1$ is defined as follows: $a \uparrow b := a^b$, $a \uparrow^2 b := a^{a^{\dots^{a}}}$ ($b$ levels), and more generally
    \[
    a\uparrow^k b :=  \begin{cases} a^b & k = 1\\ a & b=1, k \ge 2\\ a \uparrow^{k-1} (a \uparrow^{k} (b-1)) & b, k\ge 2 \end{cases}
    \]
    Note that $\min \{k \mid 3 \uparrow^k 2 > m\} = \alpha(m) \pm O(1)$. (This follows from the fact that the bivariate definition of the Ackerman function as from~\cite{Kaplan2002Union} satisfies $A(x, y) = 2 \uparrow^{x-2}(y+3) - 3$.) 
\end{definition}

\begin{figure}[hb]
    \includegraphics[]{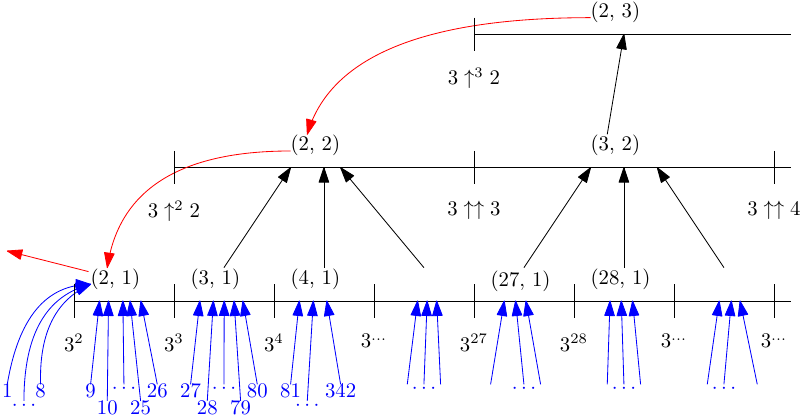}
    \caption{The tree in \cref{lemm:tree_weak}. Internal nodes are black, leaves are blue. In the proof, the red edges get added to turn the forest into a tree.}
    \label{fig:tree}
\end{figure}

To show \cref{lemm:tree_weak}, we will construct a tree as depicted in Figure~\ref{fig:tree}.
Every internal node corresponds to some subinterval of $[1, m]$,
and these nodes are arranged in levels.
More precisely, the $j$-th node on level $k$ corresponds to the half-open interval
$[3\uparrow^k j, \min(m+1, 3 \uparrow^k (j+1)))$ where $j \ge 2, k \ge 1$, and $3 \uparrow^k j \le m$.
We denote this node by $(j, k)$.
Note that the intervals of nodes on adjacent levels are nested. That is, every interval on level $k$ is the disjoint union of intervals on level $k-1$.
This defines a forest structure on the set of nodes: Nodes $(j, k)$ with $j=1$ are root nodes. All other nodes $(j, k)$ have a parent on level $k+1$.
To turn this forest into a tree, we additionally add an edge from node $(1, k+1)$ to node $(1, k)$ for $k \ge 1$.
This yields a tree with root node $(2, 1)$.
Finally, to every node $(j, 1)$, we attach $O(3^j)$ leaves, numbered $3^j, 3^j+1, \dots, \min(m, 3^{j+1}-1)$.
To the root node $(2, 1)$, we attach eight additional leaves, numbered $1, 2, \dots, 8$.

We check that the resulting tree satisfies all properties of \cref{lemm:tree}:
The tree has $m$ leaves, numbered $1, \dots, m$.
The eight leaves $1, 2, \dots, 8$ have depth $O(1)$.
For the $i$-th leaf, with $i \ge 8$, there is a unique node $(2, k)$ whose interval contains $i$.
Then $i \ge 3 \uparrow^k 2$, i.e., $k \le \alpha(i)$.
We claim that the $i$-th leaf has depth $\le 2k$.
Indeed, from the leaf, it takes at most $k$ steps to the node $(2, k)$, and another $k$ steps to the root node $(2, 1)$.
This shows property (1) of \cref{lemm:tree}.
What remains is to check property (2).
The root node has $O(1)$ children.
A node $(j, 1)$ where $j > 2$ has $O(3^j)$ children.
Since the leaves in the subtree of $(j, 1)$ are $3^j, \dots, \min(m, 3^{j+1}-1)$,
property (2) holds for these nodes.
Consider now an arbitrary node $(j, k)$ where $k \ge 2$, with interval $[3 \uparrow^k j, 3 \uparrow^k (j+1))$.
Since $3 \uparrow^k (j+1) = 3 \uparrow^{k-1}(3 \uparrow^k j)$,
the children of this node have an interval of the form $[3 \uparrow^{k-1} \ell, 3 \uparrow^{k-1} (\ell+1))$ where $\ell+1 \le 3 \uparrow^k j$.
Thus, the node $(j, k)$ has $\le 3 \uparrow^k j$ children.
On the other hand, if the $i$-th leaf lies in the subtree of the node $(j, k)$,
then $i \in [3 \uparrow^k j, 3 \uparrow^k (j+1))$, so $i$ is at least the number of children of $(j, k)$. 
Since $(j, k)$ was arbitrary, this shows property (2).

To compute this tree on a pointer machine in $O(m)$ time, we need two observations:
First, the final step of attaching leaves can be implemented in $O(m)$ time.
Second, if $3 \uparrow^k j \le m$, then we can compute $3 \uparrow^k j$ in $O(\log^3 m)$ time.
Indeed, if we use the recursion of \cref{def:up_arrow}, then there  can be at most $\log_3(m)$ multiplications in total,
and we can simulate on a pointer machine the multiplication of $O(\log m)$-bit numbers in $O(\log^2 m)$ time.
Conversely, if $3 \uparrow^k j > m$, then we can detect this in $O(\log^3(m))$ via the same approach.
Thus, we can compute the internal nodes (and their intervals) level-by-level, in $O(\log^3(m))$ time per internal node.
Since there are $O(\log m)$ internal nodes,
this step takes $o(m)$ time in total. \\

\noindent
This shows \cref{lemm:tree_weak}, which implies \cref{lemm:tree} and \cref{theo:top_level}, which concludes \cref{theo:main}.

\section{An age-aware DecreaseKey operation}

We briefly note that our result can be extended somewhat. 
In \cref{theo:main}, the \texttt{DecreaseKey} operation takes $O(\alpha(n))$ amortized time.
We can perform a slightly tighter analysis 
and show that the \texttt{DecreaseKey} operation on an element $x_j$ takes $O(\alpha(\Gamma(x_j)))$ time.

There are two steps of the \texttt{DecreaseKey} operation that take $\omega(1)$ time:
The find operation on the union-find structure, and the \texttt{DecreaseKey} operation on the index.
Since the element $x_j$ lies in a bucket of index $i \in O(\log \Gamma(x_j))$,
the \texttt{DecreaseKey} operation on the index takes $O(\alpha(i)) \subseteq O(\alpha(\Gamma(x_j)))$ time.
For the find operation, we observe that element $x_j$ lies in a set of size $O(\Gamma(x_j))$.
Thus, it suffices to prove the following:

\begin{lemma} \label{lemm:fast_find}
    There is a union-find data structure that supports make set and link operations in amortized $O(1)$ time,
    and the find operation on an element $x_j$ in amortized $O(\alpha(|S|))$ time where $S$ is the set that currently contains $x_j$.
\end{lemma}

The proof of this lemma follows immediately from refining the existing analysis of union-find, making the cost analysis more sensitive to the current set size.
Such adaptions are common~\cite{Alstrup2014Union, Kaplan2002Union}, but for completeness we provide the full argument.
We follow the analysis of \cite{tarjan1984Worst}, starting on page 255.
There, the cost of a find operation is $k$, which is later bounded by $O(\alpha(n))$.
In the proof of Lemma 3 of \cite{tarjan1984Worst}, we replace $k$ by the variable $g$ that is used in the proof.
Since $g$ is bounded by the number of distinct levels in $S$, we have $g \in O(\alpha(|S|))$.
Thus, the find operation takes amortized $O(\alpha(|S|))$ time.
This shows \cref{lemm:fast_find} and therefore:

\begin{restatable}{theorem}{main_final}
    A heap can be implemented on a pointer machine that, when $n$ denotes the heap size, supports the following operations:
    \begin{itemize}
        \item \texttt{MakeHeap} in worst-case $O(1)$ time,
        \item \texttt{DecreaseKey} in amortized $O(\alpha(\Gamma(x_j)))$ (inverse-Ackermann) time, 
        \item \texttt{Push} in amortized $O(1)$ time, 
        \item \texttt{Peek} in worst-case $O(1)$ time, and
        \item \texttt{Pop} in worst-case $O(1 + \log \Gamma(x_j))$ time.
    \end{itemize}
    All worst-case operations build no amortized potential. I.e., invoking them does not affect the amortized analysis of the other operations. 
\end{restatable}

\bibliographystyle{plainurl}
\bibliography{refs}

@inproceedings{Haeupler24,
  author       = {Haeupler, Bernhard and Hlad{\'\i}k, Richard and Rozho{\v{n}}, V{\'a}clav and Tarjan, Robert and T{\v{e}}tek, Jakub},
  title        = {Universal Optimality of Dijkstra via Beyond-Worst-Case Heaps},
  booktitle    = {Symposium on Foundations of Computer Science ({FOCS})},
  publisher    = {{IEEE}},
  year         = {2024},
  doi          = {10.1109/FOCS61266.2024.00125}
}

@article{tarjan1975efficiency,
  author  = {Robert Endre Tarjan},
  title   = {Efficiency of a Good But Not Linear Set Union Algorithm},
  journal = {Journal of the ACM},
  volume  = {22},
  number  = {2},
  pages   = {215--225},
  year    = {1975},
  doi     = {10.1145/321879.321884}
}

@inproceedings{DBLP:conf/soda/Thorup03a,
  author       = {Mikkel Thorup},
  title        = {On {AC}\({}^{\mbox{0}}\) implementations of fusion trees and atomic
                  heaps},
  booktitle    = {{ACM-SIAM} Symposium on Discrete
                  Algorithms (SODA)},
  pages        = {699--707},
  publisher    = {{ACM/SIAM}},
  year         = {2003},
  doi = {10.1145/644108.644221}
}

@article{Fredman2994Trans,
author = {Fredman, Michael L. and Willard, Dan E.},
title = {Trans-dichotomous algorithms for minimum spanning trees and shortest paths},
year = {1994},
issue_date = {June 1994},
publisher = {Academic Press, Inc.},
address = {USA},
volume = {48},
number = {3},
issn = {0022-0000},
url = {https://doi.org/10.1016/S0022-0000(05)80064-9},
doi = {10.1016/S0022-0000(05)80064-9},
journal = {J. Comput. Syst. Sci.},
month = jun,
pages = {533–551},
numpages = {19}
}

@article{fredman1993surpassing,
  title={Surpassing the information theoretic bound with fusion trees},
  author={Fredman, Michael and Willard, Dan},
  journal={Journal of computer and system sciences},
  volume={47},
  number={3},
  pages={424--436},
  year={1993},
  publisher={Elsevier},
    doi = {10.1016/0022-0000(93)90040-4}
}

@inproceedings{Kaplan2022Meld,
author = {Kaplan, Haim and Shafrir, Nira and Tarjan, Robert E.},
title = {Meldable heaps and boolean union-find},
year = {2002},
isbn = {1581134959},
publisher = {Association for Computing Machinery},
address = {New York, NY, USA},
url = {https://doi.org/10.1145/509907.509990},
doi = {10.1145/509907.509990},
booktitle = {ACM Symposium on Theory of Computing (STOC)},
pages = {573–582},
numpages = {10},
location = {Montreal, Quebec, Canada}
}

@InProceedings{Elmasry2010Violation,
author="Elmasry, Amr",
editor="Thai, My T.
and Sahni, Sartaj",
title="The Violation Heap: A Relaxed {F}ibonacci-Like Heap",
booktitle="Computing and Combinatorics",
year="2010",
publisher="Springer Berlin Heidelberg",
address="Berlin, Heidelberg",
pages="479--488",
isbn="978-3-642-14031-0"
}

@article{Haeupler2011Rank,
author = {Haeupler, Bernhard and Sen, Siddhartha and Tarjan, Robert E.},
title = {Rank-Pairing Heaps},
journal = {SIAM Journal on Computing},
volume = {40},
number = {6},
pages = {1463-1485},
year = {2011},
doi = {10.1137/100785351}
}

@Inbook{Chan2013,
author="Chan, Timothy M.",
title="Quake Heaps: A Simple Alternative to Fibonacci Heaps",
bookTitle="Space-Efficient Data Structures, Streams, and Algorithms: Papers in Honor of Ian Munro on the Occasion of His 66th Birthday",
year="2013",
publisher="Springer Berlin Heidelberg",
address="Berlin, Heidelberg",
pages="27--32",
isbn="978-3-642-40273-9",
doi="10.1007/978-3-642-40273-9_3",
url="https://doi.org/10.1007/978-3-642-40273-9_3"
}

@article{Alstrup2014Union,
author = {Alstrup, Stephen and Thorup, Mikkel and G\o{}rtz, Inge Li and Rauhe, Theis and Zwick, Uri},
title = {Union-Find with Constant Time Deletions},
year = {2014},
issue_date = {October 2014},
publisher = {Association for Computing Machinery},
address = {New York, NY, USA},
volume = {11},
number = {1},
issn = {1549-6325},
url = {https://doi.org/10.1145/2636922},
doi = {10.1145/2636922},
journal = {ACM Transactions on Algorithms},
month = aug,
articleno = {6},
numpages = {28}
}

@inproceedings{Kaplan2002Union,
author = {Kaplan, Haim and Shafrir, Nira and Tarjan, Robert E.},
title = {Union-find with deletions},
year = {2002},
isbn = {089871513X},
publisher = {Society for Industrial and Applied Mathematics},
address = {USA},
booktitle = {ACM-SIAM Symposium on Discrete Algorithms (SODA)},
pages = {19–28},
numpages = {10},
location = {San Francisco, California},
series = {SODA '02}
}

@inproceedings{iacono2000improved,
  title={Improved upper bounds for pairing heaps},
  author={Iacono, John},
  booktitle={Scandinavian Workshop on Algorithm Theory},
  pages={32--45},
  year={2000},
  organization={Springer},
    doi = {10.5555/645900.672600}
}

@article{TARJAN1979110,
title = {A class of algorithms which require nonlinear time to maintain disjoint sets},
journal = {Journal of Computer and System Sciences},
volume = {18},
number = {2},
pages = {110-127},
year = {1979},
issn = {0022-0000},
doi = {https://doi.org/10.1016/0022-0000(79)90042-4},
url = {https://www.sciencedirect.com/science/article/pii/0022000079900424},
author = {Robert Endre Tarjan}
}

@article{Vuillemin2078,
author = {Vuillemin, Jean},
title = {A data structure for manipulating priority queues},
year = {1978},
issue_date = {April 1978},
publisher = {Association for Computing Machinery},
address = {New York, NY, USA},
volume = {21},
number = {4},
issn = {0001-0782},
url = {https://doi.org/10.1145/359460.359478},
doi = {10.1145/359460.359478},
journal = {Communications of the ACM},
month = apr,
pages = {309–315},
numpages = {7}
}

@book{DBLP:books/aw/Knuth68,
  author    = {Donald E. Knuth},
  title     = {The Art of Computer Programming, Volume {I:} Fundamental Algorithms},
  publisher = {Addison-Wesley},
  year      = {1968},
  bibsource = {dblp computer science bibliography, https://dblp.org}
}

@inproceedings{Hoog2025Simpler,
  title={Simpler Optimal Sorting from a Directed Acyclic Graph},
  author={van der Hoog, Ivor and Rotenberg, Eva and Rutschmann, Daniel},
  booktitle={SIAM Symposium on Simplicity in Algorithms (SOSA)},
  year={2025},
doi = {10.1137/1.9781611978315.26}
}

@inproceedings{Pettie2005Towards,
author = {Pettie, Seth},
title = {Towards a Final Analysis of Pairing Heaps},
year = {2005},
isbn = {0769524680},
publisher = {IEEE Computer Society},
address = {USA},
url = {https://doi.org/10.1109/SFCS.2005.75},
doi = {10.1109/SFCS.2005.75},
booktitle = {Symposium on Foundations of Computer Science (FOCS)},
pages = {174–183},
numpages = {10},
series = {FOCS '05}
}

@article{Fredman1999Pairing,
author = {Fredman, Michael L.},
title = {On the efficiency of pairing heaps and related data structures},
year = {1999},
issue_date = {July 1999},
publisher = {Association for Computing Machinery},
address = {New York, NY, USA},
volume = {46},
number = {4},
issn = {0004-5411},
doi = {10.1145/320211.320214},
journal = {Journal of the ACM (JACM)},
month = jul,
pages = {473–501},
numpages = {29}
}

@article{Fredman1986Pairing,
author = {Fredman, Michael L. and Sedgewick, Robert and Sleator, Daniel D. and Tarjan, Robert E.},
title = {The pairing heap: A new form of self-adjusting heap},
year = {1986},
issue_date = {Nov 1986},
publisher = {Springer-Verlag},
address = {Berlin, Heidelberg},
volume = {1},
number = {1–4},
issn = {0178-4617},
doi = {10.1007/BF01840439},
journal = {Algorithmica},
month = jan,
pages = {111–129},
numpages = {19}
}

@article{Dijkstra1959,
  author    = {Edsger Dijkstra},
  title     = {A Note on Two Problems in Connexion with Graphs},
  journal   = {Numerische Mathematik},
  volume    = {1},
  pages     = {269--271},
  year      = {1959},
  doi       = {10.1007/BF01386390}
}

@article{fredman1987fibonacci,
  title={Fibonacci heaps and their uses in improved network optimization algorithms},
  author={Fredman, Michael and Tarjan, Robert},
  journal={Journal of the ACM (JACM)},
  volume={34},
  number={3},
  pages={596--615},
  year={1987},
  publisher={ACM New York, NY, USA},
    doi = {10.1145/28869.28874}
}

@phdthesis{iacono2001distribution,
author = {Iacono, John},
title = {Distribution-sensitive data structures},
school = {Rutgers University},
year = {2001},
isbn = {049341858X},
note = {AAI3029688}
}

@inproceedings{Iacono2001Alternatives,
  author       = {John Iacono},
  title        = {Alternatives to splay trees with O(log n) worst-case access times},
  booktitle    = { Symposium on Discrete Algorithms {SODA}},
  pages        = {516--522},
  publisher    = {{ACM/SIAM}},
  year         = {2001},
  url          = {http://dl.acm.org/citation.cfm?id=365411.365522},
}

@phdthesis{crane1972linear,
  author = {Clark Allan Crane},
  title = {Linear Lists and Priority Queues as Balanced Binary Trees},
  school = {Stanford University},
  address = {Stanford, CA, USA},
  year = {1972},
  note = {PhD Thesis}
}

@article{WilliamsHeapsort,
author = {Williams, J.W.J.},
title = {Algorithm 232: Heapsort},
year = {2025},
issue_date = {June 1964},
publisher = {Association for Computing Machinery},
address = {New York, NY, USA},
volume = {7},
number = {6},
issn = {0001-0782},
doi = {10.1145/512274.3734138},
journal = {Commun. ACM},
month = may,
pages = {347–348},
numpages = {2}
}

@book{CLRS2009,
  author    = {Thomas H. Cormen and Charles E. Leiserson and Ronald L. Rivest and Clifford Stein},
  title     = {Introduction to Algorithms},
  edition   = {3},
  publisher = {MIT Press},
  year      = {2009}
}

@article{BENTLEY1980301,
title = {Decomposable searching problems {I.} {S}tatic-to-dynamic transformation},
journal = {Journal of Algorithms},
volume = {1},
number = {4},
pages = {301-358},
year = {1980},
issn = {0196-6774},
doi = {https://doi.org/10.1016/0196-6774(80)90015-2},
url = {https://www.sciencedirect.com/science/article/pii/0196677480900152},
author = {Jon Louis Bentley and James B Saxe},
}

@InProceedings{hoog2025SimplerDijkstra,
  author =	{van der Hoog, Ivor and Rotenberg, Eva and Rutschmann, Daniel},
  title =	{{Simpler Universally Optimal Dijkstra}},
  booktitle =	{European Symposium on Algorithms (ESA)},
  pages =	{71:1--71:9},
  series =	{Leibniz International Proceedings in Informatics (LIPIcs)},
  ISBN =	{978-3-95977-395-9},
  ISSN =	{1868-8969},
  year =	{2025},
  volume =	{351},
  editor =	{Benoit, Anne and Kaplan, Haim and Wild, Sebastian and Herman, Grzegorz},
  publisher =	{Schloss Dagstuhl -- Leibniz-Zentrum f{\"u}r Informatik},
  address =	{Dagstuhl, Germany},
  doi =		{10.4230/LIPIcs.ESA.2025.71}
}

@article{Hansen2017Hallow,
author = {Hansen, Thomas Dueholm and Kaplan, Haim and Tarjan, Robert E. and Zwick, Uri},
title = {Hollow Heaps},
year = {2017},
issue_date = {July 2017},
publisher = {Association for Computing Machinery},
address = {New York, NY, USA},
volume = {13},
number = {3},
issn = {1549-6325},
doi = {10.1145/3093240},
journal = {ACM Transactions on Algorithms (TALG)},
month = jul,
articleno = {42},
numpages = {27}
}

@InProceedings{hartmann2021Analysis,
  author =	{Hartmann, Maria and Kozma, L\'{a}szl\'{o} and Sinnamon, Corwin and Tarjan, Robert E.},
  title =	{{Analysis of Smooth Heaps and Slim Heaps}},
  booktitle =	{International Colloquium on Automata, Languages, and Programming (ICALP)},
  pages =	{79:1--79:20},
  series =	{Leibniz International Proceedings in Informatics (LIPIcs)},
  ISBN =	{978-3-95977-195-5},
  ISSN =	{1868-8969},
  year =	{2021},
  volume =	{198},
  editor =	{Bansal, Nikhil and Merelli, Emanuela and Worrell, James},
  publisher =	{Schloss Dagstuhl -- Leibniz-Zentrum f{\"u}r Informatik},
  address =	{Dagstuhl, Germany},
  doi =		{10.4230/LIPIcs.ICALP.2021.79}
}

@article{KaplanTarjanZwick2014,
  author    = {Haim Kaplan and Robert E. Tarjan and Uri Zwick},
  title     = {Fibonacci Heaps Revisited},
  journal   = {arXiv preprint arXiv:1407.5750},
  year      = {2014},
  eprint    = {1407.5750},
  archivePrefix = {arXiv},
  primaryClass  = {cs.DS}
}

@article{Brodan2025Strict,
author = {Brodal, Gerth St\o{}lting and Lagogiannis, George and Tarjan, Robert E.},
title = {Strict {F}ibonacci Heaps},
year = {2025},
issue_date = {April 2025},
publisher = {Association for Computing Machinery},
address = {New York, NY, USA},
volume = {21},
number = {2},
issn = {1549-6325},
url = {https://doi.org/10.1145/3707692},
doi = {10.1145/3707692},
journal = {ACM Trans. Algorithms},
month = jan,
articleno = {15},
numpages = {18}
}

@article{tarjan1984Worst,
author = {Tarjan, Robert E. and van Leeuwen, Jan},
title = {Worst-case Analysis of Set Union Algorithms},
year = {1984},
issue_date = {April 1984},
publisher = {Association for Computing Machinery},
address = {New York, NY, USA},
volume = {31},
number = {2},
issn = {0004-5411},
url = {https://doi.org/10.1145/62.2160},
doi = {10.1145/62.2160},
journal = {J. ACM},
month = mar,
pages = {245–281},
numpages = {37}
}

@article{elmasry_priority_2012,
	series = {Selected papers from the 22nd {International} {Workshop} on {Combinatorial} {Algorithms} ({IWOCA} 2011)},
	title = {A priority queue with the time-finger property},
	volume = {16},
	issn = {1570-8667},
	url = {https://www.sciencedirect.com/science/article/pii/S1570866712000834},
	doi = {10.1016/j.jda.2012.04.014},
	urldate = {2026-03-17},
	journal = {Journal of Discrete Algorithms},
	author = {Elmasry, Amr and Farzan, Arash and Iacono, John},
	month = oct,
	year = {2012},
	pages = {206--212},
}

@article{elmasry_priority_2006,
	title = {A priority queue with the working-set property},
	volume = {17},
	issn = {0129-0541},
	url = {https://www.worldscientific.com/doi/abs/10.1142/S0129054106004510},
	doi = {10.1142/S0129054106004510},
	number = {06},
	urldate = {2026-03-27},
	journal = {International Journal of Foundations of Computer Science},
	publisher = {World Scientific Publishing Co.},
	author = {Elmasry, Amr},
	month = dec,
	year = {2006},
	pages = {1455--1465},
}

@inproceedings{kozma_smooth_2018,
	address = {New York, NY, USA},
	series = {{STOC} 2018},
	title = {Smooth heaps and a dual view of self-adjusting data structures},
	isbn = {978-1-4503-5559-9},
	url = {https://dl.acm.org/doi/10.1145/3188745.3188864},
	doi = {10.1145/3188745.3188864},
	urldate = {2026-02-18},
	booktitle = {Proceedings of the 50th {Annual} {ACM} {SIGACT} {Symposium} on {Theory} of {Computing}},
	publisher = {Association for Computing Machinery},
	author = {Kozma, László and Saranurak, Thatchaphol},
	month = jun,
	year = {2018},
	pages = {801--814},
}

@misc{rutschmann2026sorting,
      title={Sorting under Partial Information with Optimal Preprocessing Time via Unified Bound Heaps}, 
      author={Daniel Rutschmann},
      year={2026},
      eprint={2604.12653},
      archivePrefix={arXiv},
      primaryClass={cs.DS},
      url={https://arxiv.org/abs/2604.12653}, 
}

\end{document}